\documentclass[amssymb,amsmath,prb,showpacs,showkeys,superscriptaddress]{revtex4}

\usepackage{amssymb,epsfig,graphics,rotating}
\DeclareFontShape{OML}{cmm}{m}{b}{%
   <-> cmmib10}{}
\DeclareMathAlphabet{\mathbf}{OML}{cmm}{m}{b}
\DeclareSymbolFont{boldletters}{OML}{cmm}{m}{b}
\DeclareMathSymbol{\balpha}{\mathord}{boldletters}{11}
\DeclareMathSymbol{\bbeta}{\mathord}{boldletters}{12}
\DeclareMathSymbol{\bgamma}{\mathord}{boldletters}{13}
\DeclareMathSymbol{\bomega}{\mathord}{boldletters}{33}
\DeclareMathSymbol{\bsigma}{\mathord}{boldletters}{27}
\DeclareMathSymbol{\btau}{\mathord}{boldletters}{28}
\def\bm#1{\mbox{\boldmath $#1$}}

\def\I{\rm i}

\newcommand{\be}{\begin{equation}}
\newcommand{\ee}{\end{equation}}
\newcommand{\no}{\nonumber}
\newcommand{\bea}{\begin{eqnarray}}
\newcommand{\eea}{\end{eqnarray}}

\newcommand{\bS}{\mbox{\boldmath $S$}}
\def\case#1#2{\textstyle{#1\over#2}\displaystyle}
\def\i{{\rm i}}

\def\spr#1#2{\bS^{(#1)}\cdot\bS^{(#2)}}
\def\com#1#2{\left[#1,#2\right]}

\begin{document}
\title{Magnetization Plateaux in Bethe Ansatz Solvable Spin-$S$ Ladders}
\author{M. Maslen}
\affiliation{Department of Mathematics, Mathematical Sciences Institute, 
Australian National University, Canberra ACT 0200, Australia}
\author{M. T. Batchelor}
\email{Murray.Batchelor@anu.edu.au}
\affiliation{Department of Theoretical Physics, 
Research School of Physical Sciences and Engineering, 
Australian National University, Canberra ACT 0200, Australia}
\affiliation{Centre for Mathematics and its Applications, Mathematical Sciences Insitute,
Australian National University, Canberra ACT 0200, Australia}
\author{J. de Gier}
\email{jdegier@unimelb.edu.au}
\affiliation{Department of Mathematics and Statistics, University of Melbourne, 
Parkville VIC 3010, Australia}
\date{\today}
\begin{abstract}
We examine the properties of the Bethe Ansatz solvable two- and
three-leg spin-$S$ ladders.
These models include Heisenberg rung interactions of arbitrary
strength and thus capture the physics of the spin-$S$ Heisenberg
ladders for strong rung coupling. The discrete values derived for the magnetization 
plateaux are seen to fit with the general prediction based on the Lieb-Schultz-Mattis theorem.
We examine the magnetic phase diagram of the spin-1 ladder in detail
and find an extended magnetization plateau at the fractional value 
$\langle M \rangle =  \frac{1}{2}$ in agreement with the 
experimental observation for the spin-1 ladder compound BIP-TENO.
\end{abstract}
\maketitle

\bibliographystyle{apsrev}

\section{Introduction}
After some intensive effort, the properties of the spin-$\frac12$ 
Heisenberg ladder are now well understood. \cite{DR,D}
The physics of the model is driven by the rung basis states, which
dominate in the limit of strong rung coupling $J$.
The spin gap originating between the singlet and triplet rung states
for large $J$ is known to persist for all finite $J$,
vanishing only at the decoupling point $J=0$ where the model
reduces to two Heisenberg chains.
More generally, the $n$-leg spin-$\frac12$
Heisenberg ladders possess the rather striking property of being
massive for $n$ even and massless for $n$ odd.
The ladder models also possess interesting magnetic properties,
with clear evidence of magnetization plateaux. \cite{mag,CHP,CGHP}
From quite general arguments, the magnetization $M$ of the   
spin-$S$ $n$-leg ladders should satisfy the discrete values
\begin{equation}
n S (1 - M) \in {\mathbb Z}.  \label{mag}
\end{equation}
Although the Heisenberg ladders are not solvable in the sense of the
spin-$\frac12$ Heisenberg chain, a number of solvable ladder models
have been found.
These include modified models with dimer or matrix product
ground states \cite{KM,HMT,CB}
and models solved by means of the Bethe Ansatz.
Here we consider the latter class.
These models include a two-leg spin-$\frac12$ model, \cite{W} which
compared to the pure Heisenberg ladder, has an additional
four-spin interaction.
The model is massless in the absence
of rung interactions with a transition to a massive phase at a critical
(non-zero) value of the rung coupling.
In this solvable model the Heisenberg rung interactions appear as
chemical potentials which break the underlying $su(4)$ symmetry.
The phase diagram has been established by means of the
Bethe Ansatz solution. \cite{W,gb}
In particular, magnetization plateaux are found at the values
$m=0$ and $m=1$, in agreement with (\ref{mag}).
However, the Bethe Ansatz calculations also reveal an
infinitesimally thin magnetization plateaux
at $m=\frac12$, which may open up with the introduction of some
anisotropy into the system. This thin plateau may be interpreted
as a `plateau boundary'.

The solvable two-leg ladder model has been generalized to an arbitrary number $n$ of legs
in which the underlying $su(2^n)$ symmetry is broken
by the Heisenberg rung interactions, which again appear as
chemical potentials in the Bethe Ansatz solution. \cite{BM,gbm}
Solvable ladder models with Hubbard \cite{F} and $t-J$ \cite{FK}
interactions have also been found.
Yet others have been constructed with mixed spin \cite{mixed} and 
higher symmetries arising from other known integrable models. \cite{BGLM, foerster}
Other models \cite{LF} have been solved via mapping from the Hubbard
model with appropriate boundary conditions.
The $su(4)$ ladder model has also been extended to include an additional 
free parameter. \cite{tonel}
For the solvable 3-leg spin-$\frac12$ model, \cite{BM,gbm}
magnetization plateaux are found at the values $m=\frac13$ and
$m=1$ in agreement with (\ref{mag}). \cite{gb}
An additional infinitesimally thin plateau is also seen at $m=\frac12$.

In this paper we investigate the properties of the Bethe Ansatz
solvable two- and three-leg ladder models for general spin-$S$.
These models are defined in Section \ref{se:spinsmodel}.
In Sections \ref{se:plateaux} and \ref{se:boundaries} 
we look respectively at the magnetization 
plateaux and their boundaries, and in Section \ref{se:spin1eg} 
we give a detailed analysis of the two-leg spin-1 ladder.
Our work on the solvable spin-1 ladder is motivated by the recent experimental and 
theoretical work on certain organic materials.
For example, the compound BIP-TENO was studied experimentally, \cite{KHIG} and found to 
have magnetization plateaux of $0, \frac12$, and 1.  
The crystal structure of this substance may be modelled effectively by 
a two-leg spin-1 ladder. 
Some models have been used to propose mechanisms by which
the fractional plateau appears. \cite{OOS}  
Relatively little work has been done on the spin-1 Heisenberg ladder.
It has been studied in the semiclassical limit
via a mapping to the nonlinear sigma model and by quantum Monte Carlo
simulation and bosonisation techniques. \cite{Sen, AS}

Concluding remarks are given in section \ref{se:conclusion}.

\section{The models}
\label{se:spinsmodel}

We consider the integrable $n$-leg spin-$S$ ladders, or spin tubes with
periodic boundary conditions. As usual a `tube' is a ladder with an 
additional interaction between leg $n$ and leg $1$.
We restrict our attention to the cases $n=2$ and $n=3$ with uniform spins
on each leg, and not the more general mixed spin case. 
The spins along each leg and rung have an isotropic Heisenberg interaction, 
with the introduction of many-body terms to retain integrability. 
The $su(2)$ spin-$S$ operators $\bm S = (S^x,S^y,S^z)$ are defined by
\begin{equation}
S^x = \case{1}{2} (S^{+}+S^{-}),
\quad S^y = -\case{1}{2}\,\i\, (S^{+}-S^{-}),
\quad S^z = \mbox{diag} \{S,S-1,\ldots,-S\},
\end{equation}
where, for given $S$,
\begin{eqnarray} 
S^{+}_{m',m}&=&\sqrt{S(S+1)-m(m-1)} \; \delta_{m',m-1},\\
S^{-}_{m',m}&=&\sqrt{S(S+1)-m(m+1)} \; \delta_{m',m+1}.
\end{eqnarray}
and $m,m' = -S,-S+1,\ldots,S$. The Hamiltonian of the model is
\begin{equation}
H = \sum_{i=1}^L H^{\rm leg}_{i,i+1} + \sum_{i=1}^L H^{\rm rung}_i,
\label{eq:spinsHam}
\end{equation}
where $L$ is the number of rungs.
The leg term is given by 
\begin{equation}
H^{\rm leg}_{i,i+1} = P_{i,i+1}^{(2S+1)} \otimes P_{i,i+1}^{(2S+1)} \otimes 
\cdots \otimes P_{i,i+1}^{(2S+1)} \qquad \mbox{($n$ factors),}
\label{eq:hleg}
\end{equation}
where
\begin{equation}
P^{(2S+1)}_{i,j} = \sum_{\alpha=0}^{2S} (-1)^{2S-\alpha}
\prod_{\beta \neq \alpha}^{2S} \frac{{\bS}_i \cdot {\bS}_j 
- x_{\beta}}{x_{\alpha} - x_{\beta}}, \label{eq:spinscor}
\end{equation}
with $x_{\alpha} = \frac{1}{2}\alpha(\alpha+1) -S(S+1)$.  
Throughout this paper, site $L+1$ is identified with site 1.
For the simplest case of $S=\frac{1}{2}$, one recovers the familiar Heisenberg
operator.
Similarly for $S=1$ and $S=\frac32$,
\begin{eqnarray}
P^{(3)}_{i,i+1}&=& ({\bS}_{i}\cdot {\bS}_{i+1})^2 + {\bS}_{i}\cdot {\bS}_{i+1} - 1, \\
P^{(4)}_{i,i+1}&=& \case29 ({\bS}_{i}\cdot {\bS}_{i+1})^3 +
\case{11}{18} ({\bS}_{i}\cdot {\bS}_{i+1})^2 -
\case98 {\bS}_{i}\cdot {\bS}_{i+1} - \case{67}{32}\,.
\end{eqnarray}
The rung term has pure Heisenberg interactions, given by
\begin{equation}
H^{\rm rung}_i = \sum_{l=1}^{n-\epsilon} J_l \left(
\bS_i^{(l)} \cdot \bS_i^{(l+1)} -1 \right).\label{eq:hrung}
\end{equation}
where $\epsilon = 0$ for tubes and $\epsilon=1$ for ladders, and 
$\bS_i^{(n+1)} = \bS_i^{(1)}$. We take periodic boundary conditions along the ladder, 
and consider isotropic rung interactions ($J_l=J$). 
It has been shown that \cite{mixed}
$[H^{\rm leg}_{i,j}, H^{\rm rung}_i + H^{\rm rung}_{j}] =0$, 
so $H$ defined in (\ref{eq:spinsHam}) is also integrable for
sufficiently small $n$. 

A key ingredient in the models under consideration is that they
are built from permutators.
Such models can be diagonalized using the (nested) Bethe Ansatz. \cite{Suth} 
The $su(N)$ Bethe Ansatz equations are
\begin{eqnarray}
\qquad
\left(\frac{\lambda_j^{(1)}-\case{1}{2} \I}
           {\lambda_j^{(1)}+\case{1}{2} \I} \right)^L
&=&\prod_{k\neq j}^{M_1} \frac{\lambda_j^{(1)}-\lambda_k^{(1)} -\I}
                            {\lambda_j^{(1)}-\lambda_k^{(1)} +\I}
\prod_{k=1}^{M_2}      \frac{\lambda_j^{(1)}-\lambda_k^{(2)}+\case{1}{2} \I}
                            {\lambda_j^{(1)}-\lambda_k^{(2)}-\case{1}{2} \I},
\nonumber\\
\prod_{k\neq j}^{M_r} \frac{\lambda_j^{(r)}-\lambda_k^{(r)}-\I}
                           {\lambda_j^{(r)}-\lambda_k^{(r)}+\I}
&=&\prod_{k=1}^{M_{r-1}} \frac{\lambda_j^{(r)}-\lambda_k^{(r-1)}-\case{1}{2} \I}
                            {\lambda_j^{(r)}-\lambda_k^{(r-1)}+\case{1}{2} \I}
 \prod_{k=1}^{M_{r+1}} \frac{\lambda_j^{(r)}-\lambda_k^{(r+1)}-\case{1}{2} \I}
                            {\lambda_j^{(r)}-\lambda_k^{(r+1)}+\case{1}{2} \I}\,,
\end{eqnarray}
where in this case $N = (2S+1)^n$.
As usual the eigenenergies of $\sum_{i=1}^L H^{\rm leg}_{i,i+1}$ are
\bea
E=L-\sum_{j=1}^{M_1}\frac 1{\left(\lambda_j^{(1)}\right)^2+\frac14}\,.
\eea

It is important to note that since the permutator (\ref{eq:hleg}) is
invariant under any ordering of the basis states, this
result may be obtained using any choice of reference state (or
pseudo-vacuum) $|\Omega\rangle$ and any assignment of Bethe Ansatz
pseudo particles. For each choice, however, one has to re-interpret the
numbers $M_r$ in terms of those corresponding to the ordering chosen.
The rung Hamiltonian is dependent on the choice
of ordering, but the change is just a rearrangement of its eigenvalues
along the diagonal. We use this property to our advantage by doing
calculations with that choice of ordering for which the Bethe Ansatz
reference state is closest to the true groundstate of the
system.

\subsection{Two-leg ladder}

We turn our attention now to the specific case of the two-leg spin-$S$
ladder, for the moment in the absence of a magnetic field.
The  rung Hamiltonian (\ref{eq:hrung}) is given by
\be
H^{\rm rung}=J (\bS^{(1)} \cdot \bS^{(2)}-\bm{1})\,,
\ee
which can be rewritten as
\begin{equation}
H^{\rm rung}=J \left[ \frac{(\bS^{(1)} + \bS^{(2)})^2 -
(\bS^{(1)})^2 - (\bS^{(2)})^2}{2}-\bm{1} \right]\,.
\end{equation}
It is convenient to change to the basis where the square and the
$z$-component of the total spin of a given rung are diagonal. Now, the
operators $(\bS^{(1)} + \bS^{(2)})^2$, ${\bS^{(1)}}^2$,
and ${\bS^{(2)}}^2$ are all diagonal with common eigenstates. 
As a result we can derive the eigenvalues of $H^{\rm rung}$ simply by
combining the eigenvalues of these operators as in the above expression. 
From elementary quantum mechanics, the respective eigenvalues of these
operators are $T(T+1)$, $S(S+1)$, and
$S(S+1)$, where the total spin $T$ can assume the values $0,\ldots,2S$.
It follows that, for a two-leg ladder, the $(2S+1)^2$ rung
states fall into $2S+1$ multiplets of total spin $T$, with eigenvalues

\begin{equation}
\lambda_{T}=\frac{J}{2} ( T(T+1)-2S(S+1)-2 ), \qquad T=0,\ldots,2S,
\label{eq:2eigs}
\end{equation}
and respective multiplicities $m_T=2 T+1$.

It is emphasized that the Hamiltonian (\ref{eq:hleg}) has the same
form in this new basis. The rung Hamiltonian (\ref{eq:hrung}) now
becomes diagonal and is given by 
\begin{equation}
\label{eq:2ham}
H^{\rm rung}={\rm
diag}\{\lambda_0,\lambda_{1},\lambda_{1},\lambda_{1},\ldots,\underbrace{\lambda_T,
\ldots,\lambda_T}_{2T+1},\ldots,
\underbrace{\lambda_{2S},\ldots,\lambda_{2S}}_{4S+1}\}\,.
\end{equation}
In section \ref{se:spin1eg} we will look at the specific example of a
two-leg spin-1 ladder in detail. 

\subsection{Three-leg tube}

We follow a similar procedure to the two-leg ladder, rewriting the
Hamiltonian and changing the basis appropriately.  In the case of the
three-tube, the $(2S+1)^3$ states fall into $\lfloor 3S\rfloor+1$ types of
multiplet ($\lfloor x\rfloor$ denotes the integer part of $x$), with total 
spin $T$, whose eigenvalues are given by

\be
\label{eq:3eigs}
\lambda_T=\frac{J}{2} (T(T+1)-3S(S+1)-6),\\
\ee
where
\be
T=\frac{2S \left(\rm mod\, 2\right)}{2},\frac{2S \left(\rm mod  \, 2\right)}{2}+1 ,\ldots,3S.
\ee

This has clear similarity to (\ref{eq:2eigs}).  However the situation now
is somewhat more complicated.  This is because in general there are
several multiplets of given spin $T$; for example two distinct doublets
having the same energy in the absence of a magnetic field.  What follows
is an analysis of exactly how the states fall into multiplets in a
three-spin system.

Let us consider the addition of three spin-$S$ objects.  We first add
the spins at sites (1) and (3) to get a combined spin-$S_{13}$ system,
and in turn, add the spin at site (2) to this combined system to get a
state of total spin $T$.  Note that the choice of ordering of addition
affects only the basis vectors used to represent a state, and not the
actual state itself.  The order chosen here facilitates computation of
ladder eigenvalues in the next subsection.  
According to quantum mechanical rules for addition of angular momenta, we
have
\begin{equation}
\label{eq:ineq1}
0 \leq S_{13} \leq 2S,
\end{equation}
and 
\begin{equation}
\label{eq:ineq2}
\left|S_{13}-S \right| \, \leq T \, \leq S_{13}+S\,.
\end{equation}
So the number of spin-$T$ multiplets is given by the number of possible
values of $S_{13}$ which satisfy (\ref{eq:ineq1}) and (\ref{eq:ineq2}).
There are two cases to consider: 
\begin{itemize}

\item[\em{case 1}]: $T \leq S$

In this case the right inequality of (\ref{eq:ineq2}) is automatically
satisfied.  The left inequality is true when $S_{13}$ assumes values in the range defined by
(\ref{eq:ineq1}) that satisfy
\begin{equation}
\left| S_{13}-S \right| \leq T,
\end{equation}
namely
\begin{equation}
S_{13} = S-T, \,S-T+1, \ldots , S+T;
\end{equation}
and there are $2T+1$ such values.

\item[{\em case 2}]: $S \leq T$

We note that (\ref{eq:ineq1}) implies that $\left| S_{13}-S\right| \leq
S$, so the left inequality of (\ref{eq:ineq2}) is automatically
satisfied.  To make the right inequality true, $S_{13}$ must assume values in the range defined by
(\ref{eq:ineq1}) that satisfy
\begin{equation}
T \leq S_{13}+S,
\end{equation}
namely
\begin{equation}
S_{13} = T-S, \, T-S+1, \, \ldots, \,2S;
\end{equation}
and there are $3S-T+1$ such values (since $2S = T-S + (3S-T)$).
\end{itemize}
To summarize these results:
In a system of three spin-$S$ objects, the number of spin-$T$
multiplets (that is, $(2T+1)$-plets) is given by
\begin{equation}
\label{eq:Theta}
\Theta (S,T) = \left\{
\begin{array}{cll}
2T+1&& S \geq T \\
3S-T+1&& S < T \leq 3S
\end{array}
\right.
\end{equation}
Multiplets of the same type have the same energy under the tube
Hamiltonian, but are distinguishable by their value of $S_{13}$, the
possible values of which are
\begin{equation}
\label{eq:s13values}
S_{13} = \left\{
\begin{array}{lll}
S-T,\, S-T+1,\ldots, S+T && S \geq T\\
T-S,\, T-S+1,\ldots,2S && S<T \leq 3S.
\end{array}
\right.
\end{equation}
So, to compute the tube eigenvalues we use equation (\ref{eq:3eigs}), and
then use (\ref{eq:Theta}) to determine the number of multiplets which will
have the given eigenvalues.

\subsection{Three-leg ladder}

The three-leg ladder presents additional difficulty because of the broken
symmetry.  However, from (\ref{eq:hrung}) we see that the ladder and tube rung
Hamiltonians are related via
\begin{equation}
H^{\rm(ladder, rung)} = H^{\rm(tube, rung)} + J(1-\bS^{(1)} \cdot \bS^{(3)})
\end{equation}
Since $\bS^{(1)} \cdot \bS^{(3)}$ commutes with the Hamiltonians (see Appendix 
\ref{ap:commute}), their
eigenvalues are similarly related via 
\begin{equation} \lambda_{T,S_{13}}^{\rm (ladder)} =
\lambda_{T}^{\rm (tube)}+J(1-\mu_{S_{13}})
\label{eq:eigrel}
\end{equation}
where $\mu_{S_{13}}$ is the eigenvalue of $\bS^{(1)} \cdot \bS^{(3)}$ for
the
state. Up to a shift and scaling, this operator is similar to the two-leg
ladder Hamiltonian.  By symmetry, the eigenvalues of $\bS^{(1)} \cdot
\bS^{(3)}$ must be the same as those of $\bS^{(1)} \cdot \bS^{(2)}$, and
hence adaptation of (\ref{eq:2eigs}) yields
\begin{equation}
\label{eq:s13eigs}
\mu_{S_{13}} = \frac{1}{2}  S_{13}(S_{13}+1)-S(S+1)   
\qquad S_{13} = 0,1,\ldots,2S
\end{equation}
Thus, modifying the Hamiltonian from the tube to the ladder
gives rise to an energy shift ($J(1-\mu_{S_{13}})$) that depends,
ultimately, on the value of $S_{13}$ for the eigenstate (the tube and
ladder Hamiltonians have the same eigenstates).  So, to calculate the
ladder eigenvalues, we proceed as follows:  first calculate the tube
eigenvalues along with the number of different type of multiplets that
will occur using (\ref{eq:Theta}). Then, using (\ref{eq:s13values}) 
compute the range of values of $S_{13}$ that will occur for a given type
of multiplet.  From these, use (\ref{eq:s13eigs})  to calculate
$\mu_{S_{13}}$ and hence the energy shift, and finally, add these to the
original tube eigenvalues to obtain the ladder eigenvalues.

The process is perhaps better clarified through an example, and we
demonstrate for the case of the three leg spin-1 ladder. We put $J=1$ for
simplicity.  From
(\ref{eq:3eigs}) we determine that the tube eigenvalues are
$-6,-5,-3,$ and $0$, with respective multiplicities 1 (one
singlet), 9 (three triplets), 10 (two quintuplets), and 7 
(a septuplet).
The eigenvalues $\mu_{S_{13}}$ in this case are
$-2,-1,$ and $1$, so the corresponding 
shifts ($S(1-\mu_{S_{13}})$) required to 
give the tube eigenvalues are $3,2,$ and $0$.

We calculate the possible $S_{13}$ values (and hence the possible
$\mu_{S_{13}}$ values) that occur for each type of multiplet as given by
equation (\ref{eq:s13values}), and then proceed to compute the energy
shifts, and finally the ladder eigenvalues. 
The results are summarized in Table \ref{tab:eigs}. 

\renewcommand{\arraystretch}{1.5}
\begin{table}
\centering
\begin{tabular}{|l|c|c|c|c|c|c|c|}      \hline
multiplet& $T$ & quantity& $\lambda^{(tube)}$ &
$S_{13}$ from &
$\mu_{S_{13}}$ & shifts & $\lambda^{(ladder)}$ \\
\phantom{m}type&&($\Theta(S,T)$)& & (\ref{eq:s13values})\phantom{21} & && \\
\hline\hline
singlet & $0$ & 1 & $-6$ & 1 & $-1$ & 2 & $-4$ \\
\hline
triplet & $1$ & 3 & $-5$ & 0,1,2 & $-2,-1,1$ & 3,2,0 & $-2,-3,-5$ \\
\hline
quintuplet & $2$ & 2 & $-3$ & 1,2 & $-1,1$ & 2,0 & $-1,-3$ \\
\hline
septuplet & $3$ & 1 & 0 & 2 & 1 & 0 & 0 \\
\hline
\end{tabular}
\caption{Tabulated results relating tube eigenvalues and ladder eigenvalues.}
\label{tab:eigs}
\end{table}

\section{Magnetization Plateaux}
\label{se:plateaux}
We now analyze the effect of introducing a magnetic field $h$ along the
positive $z$-direction.  This causes the multiplets to split into 
individual nondegenerate states according to the $z$-component of a state's spin.  
The relevant physical effect is the existence of magnetization plateaux,
which occur when, in some region of phase space 
(defined by $J$ and $h$), the groundstate consists of rungs all having
the same magnetization.
We work with the magnetization per site, which is defined as
\be
M=\frac{1}{nLS} \sum_{i=1}^L \sum_{l=1}^n \left[\left(S^z\right)_i^{(l)}\right].
\ee
This definition is constructed in such a way that $M$ has a saturation value of 1.
As we move through phase space, passing through regions where the
system is gapless, we find that the magnetization varies
continuously.  However in regions where there is a nonzero gap, the
magnetization remains constant, hence the term `magnetization plateau'. 

The Lieb-Schultz-Mattis theorem for low-dimensional magnets gives rise to the 
necessary general condition \cite{mag,lsm}
\be
Q(S_c-m_c) \in {\mathbb Z}, 
\label{eq:lsm}
\ee
for the existence of a plateau.
Here $Q$ is the spatial period as measured by the unit cell, and $S_c$ and $m_c$
are the total spin and magnetization of the unit cell respectively.  Hence for an 
$n$-leg ladder they may be taken as $nS$ and $nm$.  In this definition, $m$ is 
not designed to have a saturation value of 1, and so we may take $m=SM$.  
For translationally invariant ladders, including those of this paper, we take $Q=1$, 
hence retrieving equation (\ref{mag}).

In the case of the two-leg ladder, we note that the introduction of a
magnetic field splits the energy levels of the states within a multiplet. 
Each state is shifted from the value given in 
(\ref{eq:2eigs}) by an amount $hS^z$,where $S^z$ is the $z$-component of its spin. 
Since the multiplet states may be characterized by their value of $S^z$, 
there is a complete loss of degeneracy upon the introduction of a magnetic field. 
For a rung of total spin $T$, the lowest energy state from each one has eigenvalue
\begin{equation}
\lambda_{T,\mbox{min}}=\frac{J}{2} [ T(T+1)-2S(S+1)-6]-hT, \qquad T=0,\ldots,2S.
\end{equation}
We show that it is possible for any of
these states to be the groundstate with a nonzero gap, subject to
appropriate choice of $J$ and $h$. 
Consider first the case where the two rung states of lowest energy 
are 
from the same multiplet. Since the magnetic energy shifts are given 
by $hS^z$, it is clear that the difference in their energy due
to the rung Hamiltonian is $h$.
The leg Hamiltonian gives a contribution of $4$, so that the energy 
gap is $\Delta=h-4$. Therefore if $h>4$, all rungs are
in the same state, and the system has a gap $\Delta$.  

In the case where the two lowest energy states are from 
different multiplets (say,
$\lambda_p$ and $\lambda_q$, where $\lambda_p > \lambda_q$), the gap is
given by 
\begin{eqnarray}
\Delta &=& \lambda_p - \lambda_q -4 \no\\
&=& \frac{J}{2} [p(p+1)-q(q+1)] - h(p-q)-4.
\end{eqnarray}
This can be guaranteed to be greater than zero for any $p$ and $q$
simply by taking 
\be
J> 2\left[\frac{h(p-q)+4}{p(p+1)-q(q+1)}\right]
\ee
in the antiferromagnetic case (with a similar result in the 
ferromagnetic case).
Hence, the lowest energy state from each multiplet has some 
corresponding region of phase space where it is the sole rung state 
occurring in the groundstate, and the ladder is gapped. The different multiplets 
take on all possible values of $S_z$ allowed under the 
restriction that 
the system consists of intrinsic spins $S$.
Consequently, it is apparent that a plateau of every allowable 
magnetization is realized somewhere in phase space.
This analysis can also be applied to the three-leg case without 
significant modification.

It is simple to check that the magnetization plateax we have found here
satisfy the requirements of (\ref{eq:lsm}).  We note that, if the gaps 
discussed above are nonzero, then all rungs are in the same groundstate.  
Hence $Q=1$, and we require that $S-m=k$, where $S$ and $m$ are the spin and
magnetization of a rung, and $k$ is an integer.  However, since $m$ can take
values ranging from $-S$ to $S$ in integer steps, it is clear that this relation will be satisfied.

\section{Plateau boundaries}
\label{se:boundaries}

It is of some interest to explore what happens on the boundary between two
magnetization plateaux.  We examine the two-leg case.  The boundary may be
defined by the line on which the groundstate of each plateau is 
degenerate.\cite{gb}  
In general, we see that the states that can have minimum
energy in the quadrant $J,h>0$ are the states from each
multiplet with largest $z$-component of the spin, namely

\begin{equation}
\label{eq:minstates}
\lambda_{T,\mbox{min}}=\frac{J}{2} [T(T+1)-2S(S+1)]-hT.
\end{equation}
It is also clear that as we steadily increase the ratio $h/J$ that
$T$ will progress in order through its possible values $0,..,2S$.  That
is, the plateau boundaries will occur between successively ordered states
as in (\ref{eq:minstates}).  Thus the equations of the plateau boundaries
are given by $\lambda_{T,{\rm min}}-\lambda_{T+1,{\rm min}}=0$ or more explicitly
\begin{equation}
h=\frac{J}{2} (T+1)  \qquad T=0,1,\ldots,2S-1.
\end{equation}
Now it is a simple matter to determine what the magnetization will be on
these lines.  As usual we normalize so that $\langle M\rangle=1$ 
if $S_z=S$ on each site.  We calculate the magnetization on
these boundary lines by averaging the two states that meet there.  Now the
state corresponding to $\lambda_{T,{\rm min}}$ must have $S_z$ as large as
possible, that is, $S_z=T$.  On boundary line $T$, then, we simply average
$T$ and $T+1$, which is spread over two sites and normalized for spin. 
That is
\begin{equation}
\langle M \rangle_T = \left(\frac{T+T+1}{2}\right)/ \,2S \quad = \quad\frac{2T+1}{4S},
\qquad
T=0,1,\ldots,2S-1.
\end{equation}
Although we would not, strictly speaking, consider this a magnetization 
plateau, it still satisfies (\ref{eq:lsm}) in some sense.  We can think of the
groundstate now as being a mixture of two types of rung.  Hence for the
two-leg ladder, we take $Q=2$ to be a unit cell.  The magnetization of the unit
cell is then $2T+1$, with the total spin being some integer between 0 and
$4S$.  Hence $2(S-m)$ as defined in (\ref{eq:lsm}) is clearly an integer.

\section{Example -- the two-leg, spin-1 ladder}
\label{se:spin1eg}

We look at the specific case of the two-leg, spin-1 ladder in the presence
of a magnetic field.  In this example the leg Hamiltonian is
\be
H^{\rm leg}_{i,j} = \left[\bS_i^{(1)}
\cdot \bS_{j}^{(1)}+\left(\bS_i^{(1)}
\cdot \bS_{j}^{(1)}\right)^2 -1 \right] \left[
\bS_i^{(2)}\cdot \bS_{j}^{(2)}+
\left(\bS_i^{(2)}\cdot \bS_{j}^{(2)}\right)^2 -1 \right],
\ee
while the rung Hamiltonian is given by
\begin{equation}
H^{\rm rung}_i = J \left(
\bS_i^{(1)} \cdot \bS_i^{(2)} 
-1\right)-h\left(\left(S^z\right)^{(1)}_i+\left(S^z\right)^{(2)}_i\right),\label{eq:hrungspin1}
\end{equation}
where
\begin{equation}
S^x = \frac{1}{\sqrt{2}}\left( \begin{array}{@{}ccc@{}} 0 & 1 & 0 \\ 1 & 0
&
1 \\ 0 & 1 & 0 \end{array}
\right), \quad
S^y = \frac{1}{\sqrt{2}}\left( \begin{array}{@{}ccc@{}} 0 & -\i & 0 \\ \i
&
0 & -\i \\ 0 & \i & 0 \end{array} \right), \quad
S^z = \left(\begin{array}{@{}ccc@{}}
1 & 0 & 0 \\ 0 & 0 & 0 \\ 0 & 0 & -1 \end{array} \right),
\end{equation}
are the spin-1 operators.
We will work in the basis where the square and the
$z$-component of the total spin of a given rung, ${\mathbf S}=
\bS^{(1)} + \bS^{(2)}$, are diagonal. It
follows that the nine states on a given rung fall into a
spin-2 quintuplet, a spin-1 triplet, and a spin-0 singlet.
Let $S=|\bS|$ and let 
$S^z$ be the $z$-component of the total spin. We denote the rung basis by 
$|S,S^z\rangle$ and it is given in terms of the $(S^z)^{(l)}$ eigenstates by
\begin{eqnarray}
|0\rangle &=& |0;0\rangle = \case{1}{\sqrt{3}}
\left( |+\rangle |-\rangle - |0\rangle |0\rangle + |-\rangle |+ \rangle \right)
 \nonumber\\
|1\rangle &=& |1;1\rangle = \case{1}{\sqrt{2}}
\left( -|+\rangle |0\rangle + |0\rangle |+\rangle \right ) \nonumber\\
|2\rangle &=& |1;0\rangle = \case{1}{\sqrt{2}}
\left( -|+\rangle |-\rangle + |-\rangle |+\rangle \right )\nonumber\\
|3\rangle &=& |1;-1\rangle = \case{1}{\sqrt{2}}  
\left( -|0\rangle |-\rangle + |-\rangle |0\rangle \right )\nonumber\\
|4\rangle &=& |2;2\rangle =|+\rangle |+\rangle \nonumber\\
|5\rangle &=& |2;1\rangle = \case{1}{\sqrt{2}}
\left( |+\rangle |0\rangle + |0\rangle |+\rangle
\right) \nonumber\\
|6\rangle &=& |2;0\rangle = \case{1}{\sqrt{6}}
\left( |+\rangle |-\rangle + 2|0\rangle |0\rangle + |-\rangle |+\rangle
\right) \nonumber\\
|7\rangle &=& |2;-1\rangle = \case{1}{\sqrt{2}}
\left( |0\rangle |-\rangle + |-\rangle |0\rangle
\right ) \nonumber\\
|8\rangle &=& |2;-2\rangle = |-\rangle |-\rangle \label{eq:basis}
\end{eqnarray}
The leg Hamiltonian (\ref{eq:hleg}) has the same form in
this new basis. The rung Hamiltonian (\ref{eq:hrungspin1})  now becomes
diagonal and is given by $H^{\rm rung}={\rm diag}
\{-3J,-2J-h,-2J,-2J+h,-2h,-h,0,h,2h\}$. 

We now consider the possible orderings of the states according to 
their energy in the presence of
the magnetic field $h$.  The energy considerations in this case are quite
simple, as there is no degeneracy.  The number of possible orderings is
now quite large because the field can cause varying amounts of energy
level splitting within a given multiplet, leading to overlaps between
them.  In fact there are 12 physically possible orderings for $h>0$.  From
now on we will restrict our attention to the first quadrant ($J>0$) which
contains six orderings.  We will list these below, computing energy gaps, and 
hence deriving the phase diagram.

In computing energy gaps for each ordering, we will use a procedure similar to that 
used in work on the $su(8)$ tube.\cite{gbm}  
Since the rung Hamiltonian is diagonal, it is a simple
matter to compute its contribution to this gap.  However, calculating the contribution
of the leg Hamiltonian is more complicated. This requires taking
the thermodynamic limit of the Bethe Ansatz equations, taking Fourier
transforms, and then solving.  The final step comes down to the evaluation of an 
integral, which gave us our final result.

This integral, and hence the energy gap, ultimately depends on the 
type of mode we need to create in order to compute the energy gap.
This
follows directly from the degeneracy of 
the groundstate. It is therefore possible to give a general form of 
this integral, which depends only on the type of mode we are creating.
This allows us to easily and quickly compute the gap contribution of
a Hamiltonian that is to be solved by the Bethe Ansatz method.

If we denote by $\alpha_k$ the integral
corresponding to creating a $\lambda^{(k)}$ excitation, the general 
result is
\begin{equation}
\alpha_k = \frac{-2}{k} \left[\gamma + \log 4 + \psi(
\frac{1}{2}+\frac{1}{k})\right],
\end{equation}
where $\gamma$ is the Euler gamma constant and $\psi(x)$ is the digamma
function. This can be simplified for a specified $k$.  Here, we will 
only need the values $\alpha_1=4$ and $\alpha_2=2 \log 2 = 1.38629\ldots$

We now proceed to analyze each ordering in turn.

\begin{itemize}

\item[1.] $h<\frac{2}{3} J$

In this case the ordering is $\{|0\rangle, |1\rangle, |2\rangle,
|3\rangle, |4\rangle, |5\rangle, |6\rangle, |7\rangle, |8\rangle \}$.  The
energy is given by $E=E_{leg}+M_1 (J-h)+(M_2 +M_3)h+ M_4(2J-3h)+
(M_5+M_6+M_7+M_8)h$, where $M_k=\sum_{r=k}^9
N_{Pr}$, where $N$ is the number of states $|j\rangle$ ocurring in an
excited state, and $P$ takes the numbers ${0,1,2,3,4,5,6,7,8}$ to the
ordering we present. (In this case, $P$ is the identity permutation, in 
the cases below, it will be some other permutation). Hence the first
energy gap is $\Delta=(-2J-h)-(-3J)-4 = J-h-4$, leading to a critical line 
$h_c=J-4$.  There is another crosssover curve on which the second
excitations, ($|2\rangle$ states)  also become massless.  Although we   
cannot calculate this exactly, we can see that it intersects the above
critical line at the point $(J,h)=(4,0)$. This curve moves as $M_1$
changes, eventually traversing regions 1. and 2. to the point $(J,h)=(2
\log{2},2 \log{2})$ on the boundary of region 3, $h=J$.

Below the line $h_c=J-4$, all rungs are in state $|0\rangle$, so the total
magnetization of the system is $\langle M\rangle=0$, as would be expected.
Above this line a finite number of rungs exist in the lowest-energy
triplet state, and the magnetization varies continuously.

\item[2.] $\frac{2}{3}J< h<J$

Now the ordering is $\{|0\rangle, |1\rangle, |2\rangle, |4\rangle,
|3\rangle, |5\rangle, |6\rangle, |7\rangle, |8\rangle \}$.  This gives the
same calculation as for case 1, so the critical line is the same.  Regions
1. and 2. become distinguishable only when considering the third or higher
excitations.

\item[3.] $J<h<\frac{3}{2}J$

Now the ordering is $\{|1\rangle, |0\rangle, |4\rangle,
|2\rangle, |5\rangle, |3\rangle, |6\rangle, |7\rangle, |8\rangle \}$.
The first energy gap is $E=-J+h-4$, giving rise to a critical
line of $h_c=J+4$.   The second excitations become massless on this line  
at $(J,h)=(8,12)$, and move to the point $(J,h)=(2 \log{2},2 \log{2})$.
That is, the same point as that calculated for region 1. 

Now above this line, the groundstate consists entirely of rungs in state
$|1 \rangle$.  As a result there is a magnetization plateau of
$\langle M \rangle=\frac{1}{2}$ here.

\item[4.] $\frac{3}{2}J<h<2J$

The ordering is $\{|1\rangle, |4\rangle, |0\rangle, |2\rangle, |5\rangle,
|3\rangle, |6\rangle, |7\rangle, |8\rangle \}$.  The gap now
is $\Delta = 2J-h-4$ and the critical curve is $h_c=2J-4$.  This
intersects the critical curve of region 3. on the boundary between them.
Again the second excitations become massless at $(J,h)=(8,12)$, and move
to the point $(J,h)=(2 \log{2},4 \log{2})$ on the line $h=2J$.

Below the line $h_c=2J-4$ the groundstate consists entirely of rungs in   
the $|1\rangle$ state, yielding a magnetization of $\langle M\rangle=\frac{1}{2}$,  
being part of the same plateau as found in region 3.

\item[5.] $2J<h<3J$

We now have an ordering of  $\{|4\rangle, |1\rangle, |0\rangle,
|5\rangle, |2\rangle, |6\rangle, |3\rangle, |7\rangle, |8\rangle \}$.  The
critical line here is $h_c=2J+4$.  The second excitations do not become
massless on this curve.  They do, however, become massless at the point
$(J,h)=(2 \log{2},4 \log{2})$.

\item[6.] $h>3J$

The ordering is $\{|4\rangle, |1\rangle, |5\rangle, |0\rangle, |2\rangle,
|6\rangle, |3\rangle, |7\rangle, |8\rangle \}$, yielding an extension of
the critical line found in case 5.  The second excitations become massless
on this line at $(J,h)=(0,4)$.

Above the line $h_c=2J+4$ in regions 4. and 5. the groundstate consists
entirely of rungs in state $|4\rangle$, and the system reaches a
magnetization plateau of $\langle M\rangle=1$.  Outside the three plateaux
found here the magnetization varies continuously.
\end{itemize}

\begin{figure}
\resizebox{3.5in}{!}{\includegraphics{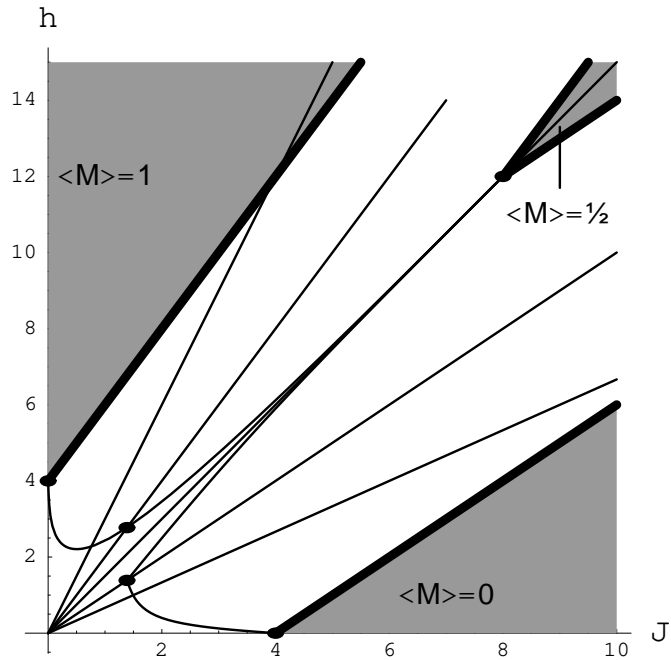}}
\caption{The magnetic phase diagram of the two-leg spin-1 ladder. The
thin lines are $h=\frac{2}{3}J$, $h=J$, $h=\frac{3}{2}J$, $h=2J$ and
$h=3J$, which together with the y-axis divide the quadrant into the  
regions 1-6. The thick lines are the exactly derived phase boundaries, and
the curved lines are sketches of higher phase boundaries.}
\label{fig:spin1phase}
\end{figure}

The results of this analysis are encapsulated in the phase diagram
(Fig.~\ref{fig:spin1phase}).  We show its first quadrant.
The shape of the 
curved line is interesting, showing a sharp spike with its cusp on the line
$h=\frac{3}{2} J$.
Although we have calculated some exact points, its overall shape cannot be
determined analytically.  We can predict some features using perturbation
theory, and it is possible that the shape of the curve could be
accurately determined using numerical series methods.  Note also that
there is more structure inside this curve, corresponding to the points at
which higher excitations become massless.

\section{Concluding remarks}
\label{se:conclusion}

We have investigated the behaviour of some Bethe Ansatz solvable two- and three-leg 
ladder models with general spin $S$.  
These models include multi-spin interactions in order to be exactly solvable.
Most importantly, the solvable ladder models include Heisenberg rung interactions of 
arbitrary strength which drive the physics for strong rung coupling. 
The values of the magnetization on each of the magnetization plateaux are seen to
fit with the general prediction based on application of the Lieb-Schultz-Mattis theorem.
Additionally, the plateau boundaries, which may be thought of as infinitesimally 
thin plateaux, also satisfy this criterion.
We calculated the phase diagram (Fig. \ref{fig:spin1phase}) of the spin-1 ladder
in detail.
Distinct plateaux occur at the values $\langle M \rangle = 0, \frac12, 1$ 
in agreement with the experimentally observed \cite{KHIG} values for the spin-1 ladder
compound BIP-TENO. 
It will be most interesting to derive the thermodynamic properties of this model.

This work has been supported by the Australian Research Council through grants
A6990558, F69700124 and DP0208925.
M. Maslen has been supported by an Australian Postgraduate Research Award.

\appendix
\section{Commutation of $\bS^{(1)}\cdot\bS^{(3)}$ with the Hamiltonians}
\label{ap:commute}
We first show that $\bS^{(1)}\cdot\bS^{(3)}$ commutes with the rung Hamiltonian. 
For the three-leg case we have
\be
H^{\rm rung} = -J ({\bm 2}-\spr12 - \spr23)\,,
\ee
so
\bea
\com{\spr13}{H^{\rm rung}} &&= -J \left(2\com{\spr13}{\bm I}-\com{\spr13}{\spr12}
-\com{\spr13}{\spr23}\right)\no\\
&&=J \left(\com{\spr13}{\spr12}
+\com{\spr13}{\spr23}\right)\no\\
&&=J \left(\com{\spr13}{\spr12+\spr23}\right)\,.\label{eq:rungcom}
\eea
This commutator may be expressed as a linear combination of commutators of the form
\be
\com{\left(S^a\right)^{(k)}\left(S^a\right)^{(l)}}
{\left(S^b\right)^{(k)}\left(S^b\right)^{(m)}+\left(S^c\right)^{(l)}\left(S^c\right)^{(m)}},
\ee
where $a, b$ and $c$ represent the co-ordinate variables $x, y$ and $z$, and $k, l, m$ 
represent the site number.  Using standard properties of commutators, 
this can be expanded to give
\be
\com{\left(S^a\right)^{(k)}}{\left(S^b\right)^{(k)}}\left(S^a\right)^{(l)}\left(S^b\right)^{(m)}
+\com{\left(S^a\right)^{(l)}}{\left(S^c\right)^{(l)}}\left(S^a\right)^{(k)}\left(S^c\right)^{(m)},
\ee
and, using the quantum mechanical relation $\com{S_j}{S_k} = {\I}\varepsilon_{jkl} S_l$, 
this becomes
\bea
&&\i\varepsilon_{abc}\left(S^c\right)^{(k)}\left(S^a\right)^{(l)}\left(S^b\right)^{(m)} + 
\i\varepsilon_{acb}\left(S^b\right)^{(l)}\left(S^a\right)^{(k)}\left(S^c\right)^{(m)}\no\\
&&=\i\left(\varepsilon_{abc}+\varepsilon_{cba}\right)\left(S^c\right)^{(k)}
\left(S^a\right)^{(l)}\left(S^b\right)^{(m)}\no\\
&&=0,
\eea
where summation is implied over repeated indices, and we have used the fact that spins from 
different sites commute.

That $\spr13$ commutes with the leg Hamiltonian follows directly from 
equation (17) of a previous paper. \cite{mixed}  
As was noted after that equation, arbitrary $XYZ$ interactions introduced
on the rungs will commute with leg Hamiltonians of the type we are working
with here.  Of course, the same statement is true for removing such 
interactions, which is how the tube Hamiltonian is transformed into the
ladder Hamiltonian. Since the Hamiltonians and the operator $\spr13$ are 
mutually commutative, the eigenvalues may be related directly as in 
(\ref{eq:eigrel}).

\bibliography{MBGfinal2}

\end{document}